# Percolation network dynamicity and sheet dynamics governed viscous behavior of poly-dispersed Graphene nano-sheet suspensions

Purbarun Dhar[1, #], Mohammad Hasan Dad Ansari[2, #], Soujit Sen Gupta[3], V. Manoj Siva[1], T. Pradeep[3], Arvind Pattamatta[1] and Sarit K. Das[1,*]

[1]Department of Mechanical Engineering, Indian Institute of Technology Madras, Chennai - 600036, India

[2]Department of Mechanical Engineering, Indian Institute of Technology Patna, Patna - 800013 India

[3]Department of Chemistry, Indian Institute of Technology Madras, Chennai - 600036, India

[#]*these authors have contributed equally to this work*

*Electronic mail: **skdas@iitm.ac.in***

## Abstract

The viscosity of Graphene nano-sheet suspensions (GNS) and its behavior with temperature and concentration have been experimentally determined. A physical mechanism for the enhanced viscosity over the base fluids has been proposed for the poly-dispersed GNSs. Experimental data reveals that enhancement of viscosity for GNSs lie in between that of Carbon Nanotube Suspensions (CNTSs) and nano-Alumina suspensions (nAS), indicating the hybrid mechanism of percolation (like CNTs) and Brownian motion assisted sheet dynamics (like Alumina particles). Sheet dynamics and percolation, along with a proposed percolation Network Dynamicity Factor; have been used to determine a dimensionally consistent analytical model to accurately determine and explain the viscosity of poly-dispersed GNSs. It has been hypothesized that the dynamic sheets behave qualitatively analogous to gas molecules. The model also provides insight into the mechanisms of viscous behavior of different dilute nanoparticle suspensions. The model has been found to be in agreement with the GNS experimental data, and even for CNT and nano-Alumina suspensions.

## 1. Introduction

Graphene, the 2-dimensional allotrope of carbon has revolutionized scientific research in the recent times. A collective ensemble of unique properties[1], it is also important to study Graphene when dispersed as micro or nano-sheets in a fluid medium. Since the inception of research in nano-suspensions or nanofluids (dilute suspensions of nanoparticles in a suitable base fluid), the academic community world over has studied their thermal properties in great details. However, in-depth research into other physical properties has remained scarce as yet. One such property is the viscosity of nano-suspensions and its implications in consequent applications. Graphene nano-suspensions (GNSs) may soon emerge as the raw materials for Graphene based thin films and printed electronic devices, asfluids with tunable electrical and/or thermal conductivities or as bio-nano-suspensions for targeted drug delivery. All such applications require motion of Graphene sheets within the fluid or of the bulk suspension itself. Thus the innate need to understand the viscosity of GNSs is of prime importance for development of the afore-mentioned technologies. In this article, we experimentally study the viscosity of GNSs and propose a mechanism for the enhancement of viscosity of the poly-dispersed GNS over the base fluid.

Studies on the viscosity of dispersed systems can be traced back to Einstein's model[2], which expressed the viscosity of the suspension as a simple linear function of the viscosity of the base fluid and the volume fraction of particle loading. However, the model has been found to be consistent with experimental data only at vanishingly low particle concentrations of spherical particles. Moreover, the model does not incorporate the effects of particle size on the viscosity of the system. Over the years, many modifications were brought about by numerous researchers onto Einstein's classical model. Some of the more famous models in this genre are those by Bachelor[3], Krieger and Dougherty[4], Eilers[5], de Bruijn[6], Pak[7] and Wang[8]. Majority of these models incorporated the effects of particle migration based on Brownian motion and added an effective higher power term of particle concentration to the existing linear model by Einstein. However, the success of the models remained confined to very low volume fractions. Also, the viscosity remained independent of the size effects and temperature effects on viscosity could not be explained by these models. The change in the viscosity of the base fluid with temperature is the only temperature-dependent term in all such models.

Several other prominent models for the viscosity of dispersed systems have been proposed based on several mechanisms, viz. Kinetic Theory[9](viscosity as an exponential function of temperature), Inter-particle spacing[10], Liquid layering[11](viscosity expressed as high order polynomial of volume fraction), particle mean free path[12, 13], Brownian diffusion of particles[14, 15], etc. Many of such models arepurely empirical in nature, lack dimensional consistency and cannot provide clear insight into the actual physical mechanism behind increase in viscosity or the viscous behavior. Furthermore, many of the models work accurately only for specific shapes and types of dispersed media and do not take into account property variations of the dispersed system. Several experimental reports on viscosity of nanofluids have been published over the last decade.Nguyen et.al[16]reported experimentalviscosity datafor $Al_2O_3$–water and CuO-water and provided completely empirical correlations for the viscosity of the nanofluids. Anoop et.al[17] looked into the electro-viscous effects in pH stabilized nanofluids and proposed a mechanism to explain the effects of particle agglomeration and effect of the Debye-Huckel screening length on the stability and viscosity of nanofluids. Yurong et.al[18] reported experimental observations for $TiO_2$ nanoparticle suspensions, however no physical explanation for the viscous behavior was provided while Namburu et.al[19] reported Newtonian behavior for CuO nanoparticles in 60:40 Ethylene Glycol and water mixture with only a deduced empirical model for nanofluid viscosity. Likewise, only experimental data and empirical correlations were provided by Kole[20] for Alumina-Engine Oil nanofluids. Viscous behavior of Chitosan stabilized Multi Wall CNT nano-suspensions were reported by Phuoc et.al[21], however, the report lacked any physical explanations for the observations. Implications of nanofluid viscosity for thermal applications were experimentally determined by Prasher et.al[22] and Murshed et.al[23], but no physical mechanism for the viscous behavior of nanofluids were provided. Also to be noted, in most of the cases discussed above, the nano-suspensions utilized were of typically high concentrations (> 2 vol.%). This leads to particle over-crowding within the fluid matrix and various other governing mechanisms creep into the system, and hence, it is recommended to study dilute systems (< 1 vol.%) to develop the clear picture. As for the present, a proper understanding of the underlying physics that govern GNS viscosity (detailed studies on the subject matter have been very rare) is needed for prolonged technical advancements in the field.

## 2. Materials and Methods

### 2.1. Nano-suspensions: Preparation and Characterization

Preparation of Graphene involves a two-step process[24]: oxidation of graphite powder to Graphite Oxide (GO) based on modified Hummer's process[25] followed by reduction of GO to Reduced GO (RGO). First, 1 g of Graphite powder (procured from R. K. Scientific Pvt. Ltd.) was taken and 12 mL of concentrated $H_2SO_4$ was added and the mixture was allowed to stand at 90°C for an hour. To it 2 grams of $K_2S_2O_8$ and $P_2O_5$ each was added with constant stirring and kept for 6 hours at the same temperature for pre-oxidation of Graphite. The mixture was then cooled to room temperature and filtered. The filtrate was discarded and the pre-oxidized GO was kept for drying in hot air over for overnight. To the dry pre oxidized GO, 24 mL of concentrated $H_2SO_4$ was added and kept at ice cold condition. Then 3 g of $KMnO_4$ was added slowly with constant stirring. The mixture was allowed to stand for 6 h under stirring condition. Then 400 mL of distilled water was added slowly under stirring condition and kept at room temperature for an hour. The reaction was stopped by adding 5 mL of 20% $H_2O_2$ and the mixture was kept undisturbed for overnight. A bright yellow colored precipitate confirmed the conversion to GO. The solution was decanted and the precipitate was washed with 1% HCl thrice. This solution was centrifuged to collect GO. The GO was dried in vacuum for 24 h. After drying, 1 g of GO was weighed and re-dispersed in De-ionized water such that the concentration of the GO solution was 0.5 wt%. This solution was further dialyzed to remove the unwanted ions. This GO solution was kept as the stock solution for preparation of Graphene. From this, 100 mL of GO solution was taken and 200 mg of NaBH4 was added and kept for stirring about an hour for reduction of the functional groups present. The solution was filtered and pre-reduced Graphene was separated and then 100 mL of water was added. This was further sulphonated with Sulphanilic acid to make Graphene nano-sheet suspensions dissolved in water as the base medium. The 0.5 wt. %RGO solution prepared was kept as the stock solution. This solution was diluted with water for getting other RGO nanofluids of lower concentrations.

The prepared Graphene was characterized by analyzing its Raman spectrum. The characteristic peak of Graphene at 1348 and 1598 cm$^{-1}$ represents the D and G band, respectively (shown in Fig. 1). The D band represents the defects and the G band represents the in-plane stretching of sp2 carbon in Graphene. The presence of 2D band at around 2800 cm$^{-1}$ characterizes Graphene formed during the chemical process. The ratio of $I_{2D}/I_G$ is around 0.34-0.5, which indicates the graphene prepared was bi/tri layered thick (Nanoscale , 2013, **5**, 381-389) For further characterization, TEM was taken (Fig. 1 inset) and the wrinkles present (marked in white arrows) confirms micro-nano scale Graphene sheets.

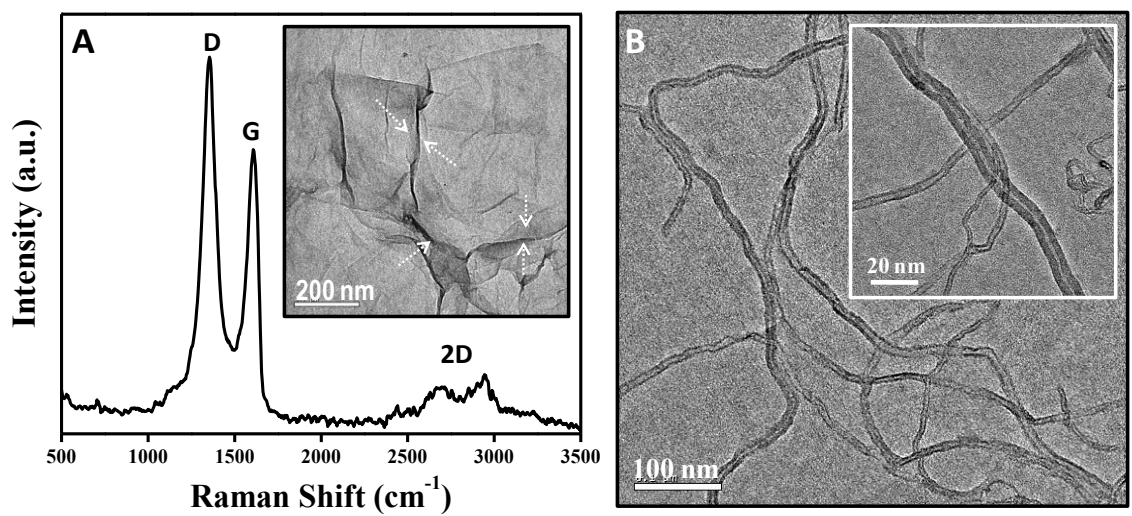

FIG1: Raman spectrum of Graphene. Inset:TEM image of Graphene

(Wrinkles inGraphene sheets have been marked with arrows).

Nano-Alumina Suspensions (nAS) and Carbon Nanotube Suspensions (CNTS) were also prepared to observe their viscous behaviors for the purpose of comparison. The method to prepare nAS involves dispersing Alumina nanoparticles as dry powder in base fluids and stabilizing the system. Alumina nanoparticles (particle size range 40-50 nm) were procured from the commercial manufacturer Alfa Aeser, USA. Required amounts of Alumina were weighted and dispersed in DI-water samples without the use of any surfactants. Ultra-sonication for 1 hour was needed to stabilize the nanofluids. The shelf life stability for such Alumina nanofluids (nAS) was found to exceed a month.

The CNTSs were similarly prepared. Required amount of dry CNT (procured from NaBond Technologies, China) was weighed and mixed in DI-water samples to get the required concentrations. The CNTSs were found to be unstable even after the ultra-sonication for two hours. Following this, Sodium dodecyl Sulfate (SDS) was used as surfactant to enhance the stability and the CNTSs were ultra-sonicated for an hour. The SDS stabilized CNTSs were found to have shelf life stability over 6 months.

## 2.2. Experimental and Measurement Details

The fluid viscosity for different loading concentrations and temperatures has been determined using an automated micro-viscometer (Anton Paar GmbH, Austria). It is a rolling-ball viscometer which works in accordance to the principle of a falling ball within a fluid media. A stainless-steel ball descends through a closed, liquid-filled capillary which is inclined at a defined angle. Inductive sensors determine the ball's time of descent. Both the dynamic and kinematic viscosity of the liquid can be calculated from the rolling time. In the present study a capillary internal diameter of 1.6mm and a steel ball of diameter 1.5mm have been used. An in-built heating element and a precision Peltier thermostat allow performing measurements at different temperatures. The angle of tilt for the capillary was set to 30° so as to allow at least 10 second descent time for all the measurements (to prevent formation of any turbulence within the capillary). Ten viscosity readings were taken for each temperature and their average was taken as the final value.

## 3. Results and Discussions:
## 3.1. Experimental Observations

The effects of temperature on the viscosity of dilute nano-suspensions were experimentally investigated. Low concentrations were studied so as to reveal the exact mechanism of enhancement in viscosity without factors like particle agglomeration, particle crowding, and sedimentation creeping into the forefront. The viscosity of the nano-suspensions has been found to increase with concentration and decrease with temperature. However, different particles have been seen to behave very differently in the fluid medium. While in the case of nAS, the effect of concentration (within dilute limits) on the viscosity of the nanofluid is negligible, the response shown by the CNTSs to concentration is highly pronounced. Interestingly enough, the trend shown by the GNSs lies nearly midway between the nAS and the CNTSs. The trends in nanofluid viscosity with concentrations at 298 K have been provided in Fig. 2. This trend in GNSs behavior has also been seen in case of GNSs thermal conductivity[26]. From such behavior, it can be hypothesized that the mechanism of viscosity enhancement in GNSs is also a hybrid of the mechanisms for nAS and CNTSs.

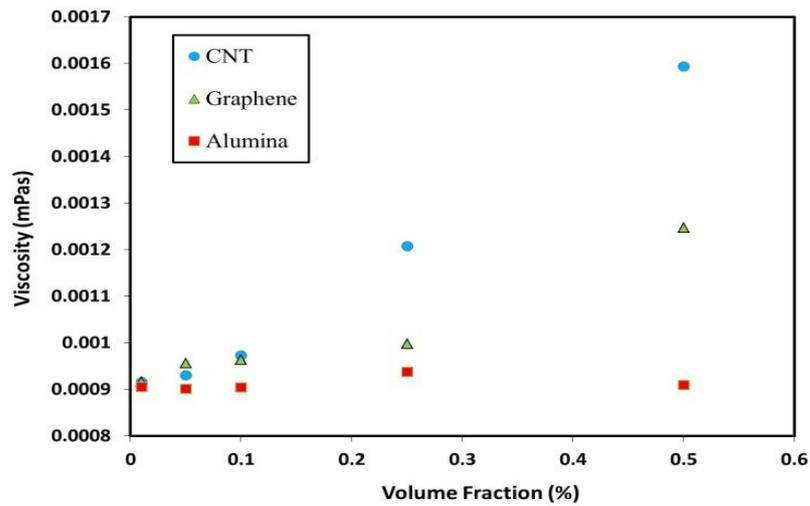

FIG. 2: Response of nanofluid viscosity to particle concentration at constant temperature (298 K)

**3.2 Mathematical Modeling**

The mechanism behind the viscous behavior of the GNSs can be explained by considering the GNS to be a poly-dispersed system, where the behavior aspects of any individual graphene sheet is dependent on the Critical sheetsize (CSS)[27] of graphene sheets for the system. The CSS theory states that for each dispersed media- base fluid pair; there exists a critical particle or sheet size that determines particle behavior within the suspension. Particles of sizes larger than this will show tendency to percolate into networks and those smaller than the critical size are dominated by the Brownian randomness within the medium. Based on this critical size, the graphene sheets can be categorized into two types, viz. micron-scale sheets, which promote viscosity by forming percolation networks, and nano-scale sheets, which promote viscosity due to their Brownian motion induced randomness. In essence, the formulation for the viscosity induced by sheet percolation is similar to that of Einstein's formulation; however, the present model is able to explain the viscous effects due to the size of the sheets, the concentration, as well as the configuration of the percolation network with temperature. The viscosity due to sheet dynamics has been derived based on dimensional analysis and the present approach provides insight into the micro-nano scale dispersed phase-fluid interactions that govern the viscosity of nano-suspensions.

The expression for the viscosity of GNSsinduced by sheet percolation has been theorized as

$$\mu_{perc} = \mu_{bf}(1 + L^* d \phi \alpha) \quad (1)$$

where $\mu_{perc}$is the viscosity of the GNS induced by the percolation networks, $\mu_{bf}$is the viscosity of the base fluid at the temperature at which the viscosity of the nano-suspensions needs to be determined, L*is the non-dimensionalized length scale and is the ratio of the average Graphene sheet face length to the CSS at that temperature ( L* = $L_G/L_{crit}$, where $L_G$ is the average sheet face size for Graphene sample and $L_{crit}$is the CSS for the Graphene –base fluid pair), d is the Percolation Network Dynamicity Factor, φ is the loading fraction (concentration) of nanoparticles and α is the Sheet distribution fraction[27]. The expression has been derived along the lines of Einstein's model[2] for viscosity of very dilute suspensions. However, it has been proposed that the viscosity of the nano-suspensions is related to the volume fraction of loading by a temperature dependent factor termed the percolation Network Dynamicity Factor (NDF).Within GNSs, the relative motion between adjacent fluid layers or between the fluid and a foreign object is hampered by the presence of the long percolation networks. It has been hypothesized that although the sheets forming these percolation networks are too massive to be affected appreciably by Brownian motion, they undergo constant rearrangements within the network itself (without compromising the integrity of network as a whole), due to the constant bombardment by the smaller Brownian dominated sheets and due to vibrations of neighboring fluid molecules.

Unlike thermal conductivity, where the mechanism is dependent on the length of the percolation paths[27], viscosity of the GNS is affected by the reshufflings occurring within the networks. This is due to the fact that viscosity of the fluid is sensitive to any relative motion with respect to the fluid molecules. Furthermore, interlayer shear within the fluid domain is higher in case of the presence of a static body than a dynamic body. As a result, static sheets within the fluid lead to more induced viscosity than dynamic sheets.Since the Brownian velocity of the smaller sheets and the vibrational energy of the fluid molecules increase with temperature, the NDF also changes. The magnitude of the NDF provides a qualitative estimate of the degree of dynamicity within the network. A smaller value of the NDF implies less reshuffling within the network and hence more stability of the network. Its behavior with temperature has been shown in Eq.(2)

$$d = A^* T^2 + B^* T + C^* \quad (2)$$

A*, B* and C* are non-adjustable constants for a particular solute-solvent pair. For Graphene-Water pair, the values of A*, B* and C* are deduced to be approximately 0.75 $K^{-2}$, -464 $K^{-1}$ and 71682 and the values have been found to be highly consistent for all volume fractions. The seemingly odd quadratic behavior of the NDF with absolute temperature (Fig. 3(d)) can be explained based on the philosophy of multiple body interactions. Qualitative illustrations have been provided in Figure 3(a), 3(b) and 3(c) for establishing the NDF hypothesis. A system composed of multiple similar components, when agitated from its initial state (Fig. 3(a)) with small agitations, will slowly begin to assemble towards a more composed state (Fig. 3(b)), i.e. a state of least total energy. However, after this if the agitations are continued with larger agitation amplitude, the state of least energy collapses and the system starts to deviate towards its original state, and with increasing amplitudes the randomness keeps increasing out of proportions (Fig. 3(c)). An example of such a system would be some sand on a tray. When the tray is slowly shaken, the sand particles will tend to collect towards one region of the tray, creating a state of minimum randomness. However, when the shaking force increases and becomes aggressive, the sand particles scatter out in all directions, increasing the randomness largely.

In case of percolation networks within GNS, temperature induced Brownian motion of the smaller sheets and the vibration of the liquid molecules act as the source of agitations. As temperature increases, the initial configurations are reshuffled and tend to make the network compact, as suggested by the decrease in the NDF. But as temperature becomes higher, the agitation increases, causing the frequency of reshuffling to increase. This leads to increased values of the NDF. Given the micron-scale size of percolating Graphene sheets (nearly equal to the CSS); the dynamicity induced becomes very high at higher temperatures.

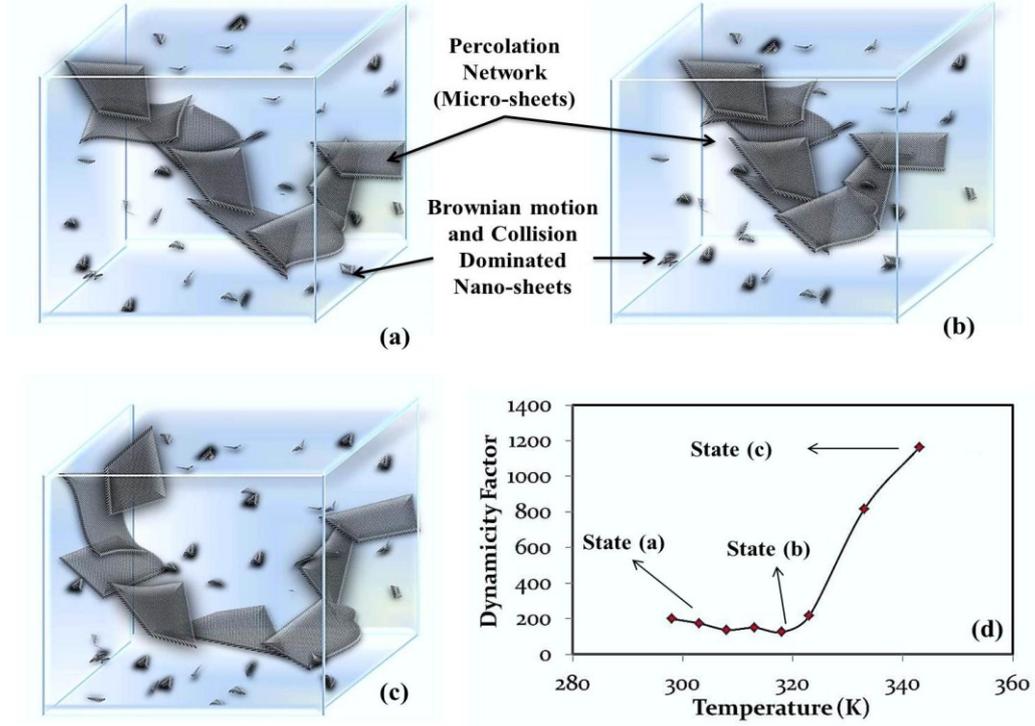

FIG.3: NDF behavior with temperature. Network configurations at (a) 300 K (b) 320 K (c) 333 K (d) the trend of NDF with temperature. The graphical configuration details for the hypothesis at different temperatures provided are purely qualitative in nature.

At first glance, the viscosity due to percolation networks seems independent of the size of individual sheets. However, the model is not completely independent of the sheet length scales. The factor L* governs the viscous behavior of the GNS. For values of L* below unity, the viscosity rendered by percolation disappears because percolation does not take place in such conditions. As the value of L* grows, the viscous effects rendered multiplies alongside it. It has been hypothesized that the viscosity due to percolation is a simple direct product of L* and the NDF.

The viscosity rendered to the GNS by sheet dynamics has been hypothesized to be qualitatively analogous to behavior of gas molecules and has been derived from dimensional analysis. The equation for the viscosity due to sheet dynamics is expressed in Eq. (3)

$$\mu_{sd} = \mu_0 \phi (1-\alpha) \qquad (3)$$

where $\mu_{sd}$ is the viscosity induced by sheet dynamics. The variable '$\mu_0$' is the dynamic viscosity term and can be expanded as

$$\mu_0 = \rho_G \lambda U_B \theta \qquad (4)$$

where, $\rho_G$ is the density of Graphene, $\lambda$ is the mean free path of inter-sheet collisions, $U_B$ is the Brownian velocity of the sheets as given by Stokes-Einstein's formula as given in Eq. (5) ($k_B$ is the Boltzmann constant, T is the absolute temperature, $L_G$ is the average face size of the dynamic Graphene nano-sheets) and $\theta$ is the collision cross-section.

$$U_B = \frac{2k_B T}{\pi \mu_{bf} L_G^2} \qquad (5)$$

The viscosity of GNS due to sheet dynamics behaves erratically and seemingly inconsistently at first sight. Among the constituent terms of '$\mu_0$' in Eq. (4), the density and mean free path terms behave in manners that lead to infer that a system with more dynamicity will have lower viscosity. On the other hand, the Brownian velocity and collision terms behaviors suggest that a dynamic system will lead to increases viscosity. The net viscosity is a resultant of such combined effects and the physical explanations for the effects have been discussedbelow.

The viscosity of the GNS due to sheet dynamics is directly proportional to the density of the dispersed media. For two particles of the same volume, the particle with higher density will be more massive and thus have a higher resistance to the momentum imparted by Brownian motion assisted inter-particle collisions. Therefore, dispersed media with higher density is more liable to be less dynamic than a dispersed system with lower density and as discussed before, a more static system renders more viscous resistance in between adjacent fluid layers. Similarly, the system will have higher induced viscosity if the mean free path of inter-particle collisions is higher. The mean free path for nanoparticles suspended in a fluid medium has been hypothesized to be of the order of a few microns at 298K. The behavior of the mean free path can be expressed as $\lambda=(L\times10^{-6})$ meter, where 'L' is a temperature dependent variant. Since it is difficult to ascertain the theoretical value of 'L' for each temperature value, the product of '$L\theta$' can be utilized (as expressed in Eqn. (6)) for ease of mathematical manipulation. The equation (Eqn. (6)) has been proposed along the lines of the assumption that the nano-sheets behave analogous to gas molecules, and hence, their mean free path of collision has been analogously considered as an exponentially decaying function of absolute temperature. A larger mean free path physically signifies low frequency of inter-particle collisions and thereby a system with lower dynamicity (and hence more viscosity) than a system with a lower mean free path (for the same volume concentration and temperature).

$$L\theta = Be^{-CT} \qquad (6)$$

The viscosity is also high for larger values of Brownian velocity. While this sounds contradictory to the explanation for mean free path, it is physically consistent. A higher value of Brownian velocity (signifying more dynamicity and hence less induced viscosity as per the explanation for mean free path) induces higher amounts of micro-scale eddies within the fluid matrix. The presence of such eddies leads to drastically increased viscosity of the system. The collision cross-section gives an idea into the probable number of inter-sheet collisions taking place within the domain at a given instant of time. With increasing number of collisions in a frozen time frame, the viscosity of the system increases. This is due to the fact that increasing number of collisions in a frozen instant of time leads to generation of more micro-scale eddies and localized micro-vortices within the system, thereby increasing the system's effective viscosity. The collision cross-section is deduced from experimental data to be a linear function of temperature and can be expressed as $\theta=(aT-b)^{27}$, where 'a' and 'b' are constants.

From the above discussion it can be concluded that unlike the viscosity rendered by sheet percolation, which can be completely eliminated by utilizing Graphene sample of sheet sizes below the CSS, the viscosity rendered by sheet dynamics becomes difficult to reduce. While the use of sheet sizes well less than the CSS leads to increased viscosity due to increased Brownian velocity and consequent micro-eddy and/or micro-vortices formation, sheet sizes closer to the CSS lead to increased viscosity due to increased shear resistance between adjacent fluid layers caused by near static behavior.

The effective viscosity of the GNS is theorized to be the summation of the two contributing viscosities and can be expressed as

$$\mu_{GNF} = \mu_{perc} + \mu_{sd} \qquad (7)$$

The viscosity of GNSs with volume concentration and temperature can be seen in Fig.(4) and Fig.(5) respectively. In the present study, the average Graphene sheet size larger than the CSS is approximately 1.5 microns (from DLS analysis) and so the value of L* is approximately 1.25. For Graphene-water pair, the values of 'B' and 'C' are determined to be $6.95 \times 10^6$ and 0.0327 respectively. The values of 'a' and 'b' are found to be 50 and 15100 respectively (for T ≥ 303 K)[27] and 1.66 and 454 respectively (for T < 303 K). In case of water as base fluid, it is observed that the number of collisions at any instant increase at a much higher rate over 303 K. This phenomenon has been found to be consistent even for nASs.

It can be seen that the model slowly deviates from the experimental values as the concentration reaches 0.5 vol. %. This is due to the fact that around and above this loading, several other factors begin to disrupt the mechanisms of network dynamicity and sheet dynamics that govern the present model. The free rearrangements and dynamicity of the networks and the sheets respectively are hampered by excessive sheet crowding (high population density of sheets within the fluid matrix) due to high concentrations. Interestingly; in such a scenario, analysis yields that the simple percolation term of the composite viscosity model can predict the viscosity at low temperatures (Fig. 5). This is evidence for excessive static behavior due to sheet crowding. However, as temperatures rise, dynamicity is gained by both the networks and the nano-sheets to some extent, and the composite model can predict the viscosity at high temperatures. This behavior is evidence that the present explanation is physically consistent.

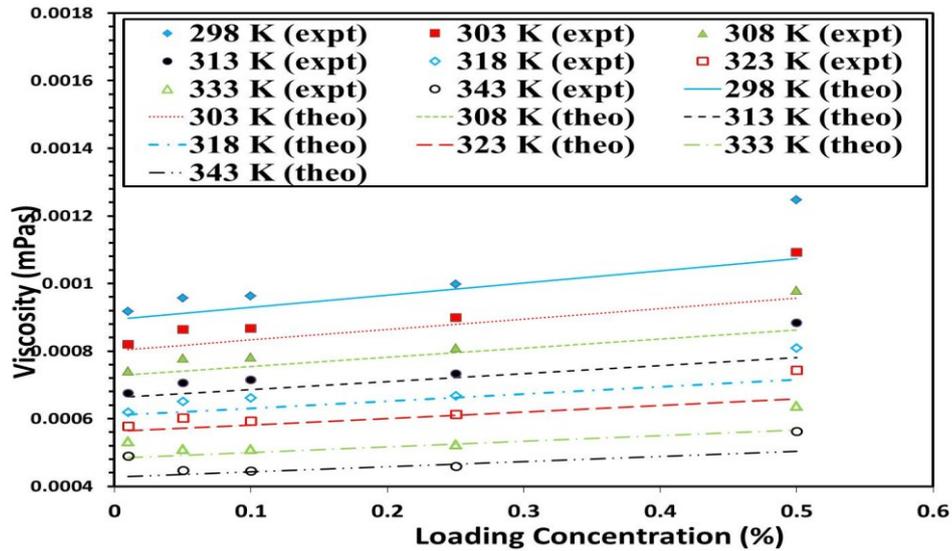

FIG. 4: Validation of analytical model for GNSwith experimental

data for different loading concentrations.

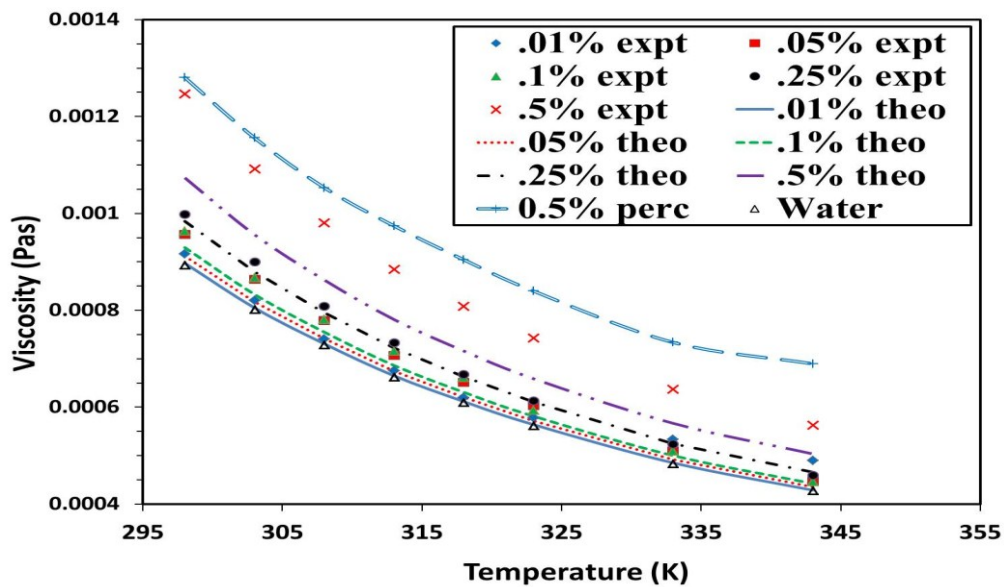

FIG.5: Validation of analytical model for GNSwith experimental

data for different temperatures.

The composite model for GNSs has been validated for a completely static system and a completely dynamic system. For a purely static system, CNT-water suspension has been used while nano-Alumina-water suspension (nAS) is used for studying purely dynamic systems. For a purely static system, i.e. CNTS, 'α' is unity and the expression for viscosity reduces to

$$\mu_{CNTNF} = \mu_{bf}(1 + L^* d\phi) \quad (8)$$

In case of CNTS, the fluid matrix is composed solely of dispersed CNT forming dense percolation networks. As explained earlier, from around 0.5 vol. %, the viscosity begins to diverge away from the predicted values. In experimental setups where the viscosity is measured by the falling ball principle within a capillary, this phenomenon is enhanced as overcrowded networks within a confined capillary provide higher viscous resistance. In the present study, CNT used has a diameter of 20 nm and length of 10 microns on an average, making the L* value approximately 10. The values of A*, B* and C* for CNT-water pair are determined to be 0.0021 $K^{-2}$, 1.377 $K^{-1}$ and 235 respectively and the behavior of the NDF with temperature can be seen in Fig. 3(b). Technically, CNTs, being of longer dimensions than Graphene sheets, should form stronger networks which exhibit much higher resilience to temperature. This fact is verified from the NDF behaviors of the two forms of carbon. While CNT networks are very less disturbed by temperature effects, the NDF of Graphene sheets are very high, implying high degrees of re-adjustment and reshuffling among the members within the networks. The plots for viscosity of CNTSs with concentration and temperature can be seen in Figs. (6) and (7) respectively.

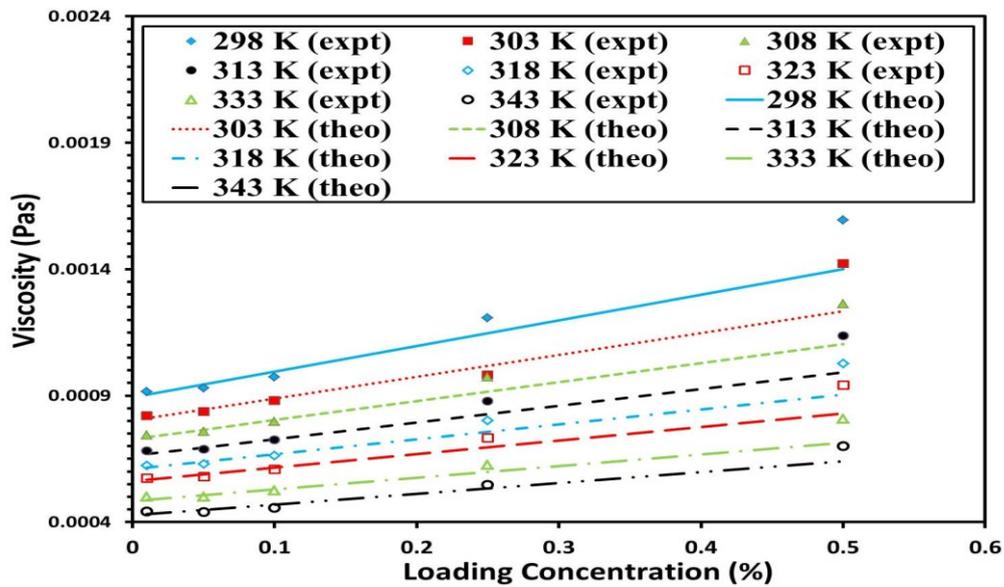

FIG. 6: Validation of analytical model for CNTS with experimental

data for different loading concentrations

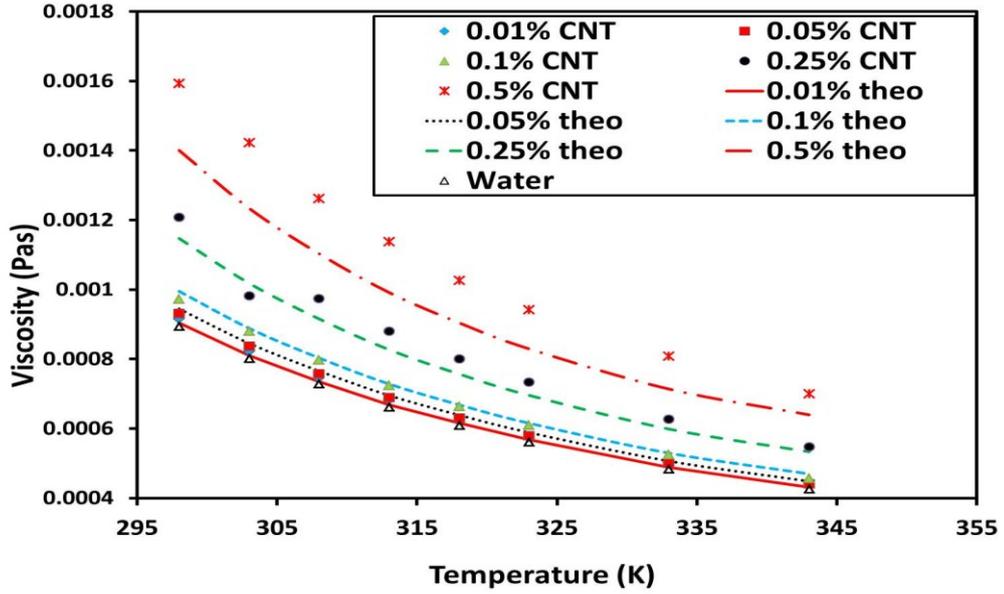

FIG. 7: Validation of analytical model for CNTS with experimental

data for different temperatures.

In case of a completely dynamic system, i.e. nAS, 'α' is zero. In such cases, the composite model collapses to simply the viscosity term rendered by particle dynamics. In case of nASs, the value of 'Lθ' is determined to be one-third that of the 'Lθ' for Graphene-water. The values for 'a' and 'b' are determined to be 15 and 4530 respectively (for T ≥ 303 K) and 0.4 and 109 respectively (for T < 303 K). The plot for nAS viscosity with temperature and concentration has been provided below.

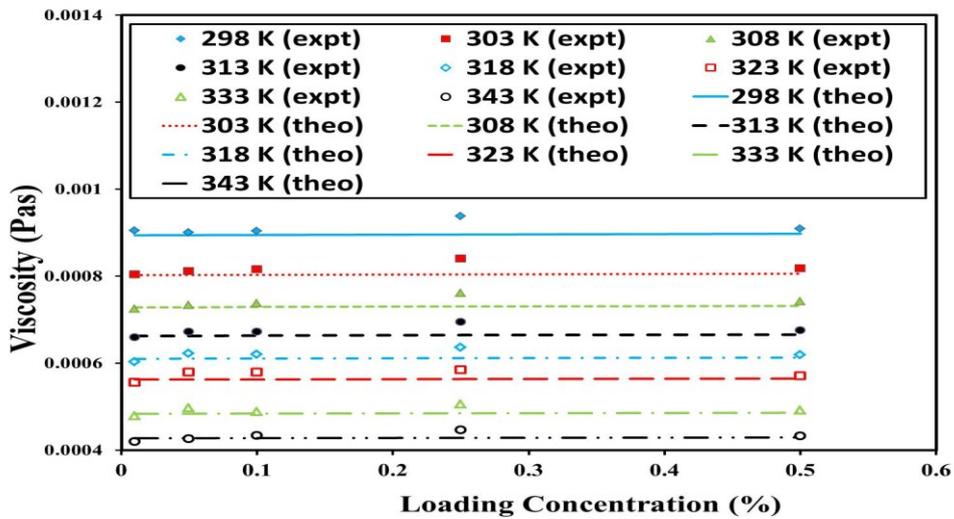

FIG. 8: Validation of analytical model for nAS with experimental

data for different loading concentrations

## 4. Conclusion

The viscosity of dilute GNSs has been experimentally determined and their response to temperature and concentration has been studied. Similar experimental studies have been carried out with CNTSs and nASs to understand the type of response exhibited by viscosity of nanofluids containing different categories of dispersed media. Viscosity response of Graphene (with its sheet like structure) suspensions were seen to behave midway between that of CNT (cylindrical geometry) and Alumina (spherical particles) suspensions. A mechanism based on this hybrid behavior and the CSS has been proposed to explain the viscous behavior of GNSs. It has been suggested that sheet percolation networks induce higher viscosity to suspensions than dynamic particles. This is evident from the viscous nature of CNTSs and nASs. The mechanism of viscosity induced by percolation networks (formed by micron scale sheets) has been hypothesized to be the temperature induced reshuffling and redistribution of individual sheets within the networks. In case of the dynamic nano-sheets, it has been suggested that they behave analogous to gas molecules within the base fluid matrix. However, their contribution to particle induced viscosity is comparatively scarce to percolation networks. The composite model for GNS viscosity has been found to be consistent and in agreement with experimental results within dilute limits. Similar consistence and agreeability has been observed in case of completely static (CNTSs) and dynamics (nASs) systems too.

To infer, the present study throws light onto the possible mechanism of solute induced viscosity in Graphene suspensions and the viscous response to temperature and concentration. As technologies where GNSs might act as the starting material, such as Graphene thin films, graphene printed electronics, graphene bio-nanofluid mediated drug delivery etc., emerge into the forefront, the importance to understand the physics behind viscosity increases, since the suitability of the base fluid for such technologies are dependent on the viscous behavior of the nanofluid.It is evident from the present study that research into newer methods to control graphene sheet sizes during preparation in liquid medium will lead to economical and precise control over GNS viscosity and the consequential applications.